\documentclass[3p,times,procedia]{elsarticle}
\usepackage[T1]{fontenc}
\usepackage[latin9]{inputenc}

\usepackage{mathrsfs}
\usepackage{amsbsy}
\usepackage{amssymb}

\usepackage{nupha_ecrc}
\usepackage{graphicx}
\usepackage{esint}

\usepackage[    bookmarks,
                 bookmarksopen = true,
                 bookmarksnumbered = true,
                 linktocpage,
                 colorlinks = true,
                 linkcolor = blue,
                 urlcolor  = blue,
                 citecolor = purple,
                 anchorcolor = green,
                 hyperindex = true,
                 hyperfigures]
                 {hyperref}

\volume{00}

\firstpage{1}

\journalname{Nuclear Physics A}

\runauth{Qun Wang}

\jid{nupha}

\jnltitlelogo{Nuclear Physics A}

\usepackage{amssymb}

\biboptions{comma,square,compress}

\usepackage[figuresright]{rotating}

\begin{document}

\begin{frontmatter}




\title{Global and local spin polarization in heavy ion collisions: a brief
overview}

\author{Qun Wang}

\address{Department of Modern Physics, University of Science and Technology
of China, Hefei, Anhui 230026, China}

\begin{abstract}
We give a brief overview about recent developments in theories and
experiments on the global and local spin polarization in heavy ion
collisions.  
\end{abstract}


\end{frontmatter}

\section{Introduction}

There is inherent correlation between rotation and polarization in
materials as shown in the Barnett effect \cite{Barnett:1935} and 
the Einstein-de Haas effect \cite{dehaas:1915}. 
We expect that the same phenomena also exist in heavy ion
collisions. Huge global angular momenta are generated in non-central
heavy ion collisions at high energies \cite{Liang:2004ph,Liang:2004xn,Voloshin:2004ha,Betz2007,Becattini:2007sr,Gao2008}.
How such huge global angular momenta are transferred to the hot and
dense matter created in heavy ion collisions and how to measure them
are two core questions in this field. There are some models to address
the first question: the microscopic spin-orbital coupling model \cite{Liang:2004ph,Liang:2004xn,Gao2008,Chen:2008wh},
the statistical-hydro model \cite{weert:1982,zubarev:1979,Becattini:2009wh,Becattini:2012tc,Becattini:2013fla,Becattini:2015nva,Hayata:2015lga}
and the kinetic model with Wigner functions \cite{Gao:2012ix,Chen:2012ca,Fang:2016vpj,Fang:2016uds}.
For the second question, it was proposed that the global angular momentum
can lead to the local polarization of hadrons, which can be measured
by the polarization of $\Lambda$ hyperons and vector mesons \cite{Liang:2004ph,Liang:2004xn}. 

The global polarization is the net polarization of local ones in an
event which is aligned in the direction of the event plane. Recently
the STAR collaboration has measured the global polarization of $\Lambda$
hyperons in the beam energy scan program \cite{STAR:2017ckg,Abelev:2007zk}.
At all energies below 62.4 GeV, positive polarizations have been found
for $\Lambda$ and $\bar{\Lambda}$. On average over all data, the
gobal polarization for $\Lambda$ and $\bar{\Lambda}$ are $\Pi_{\Lambda}=(1.08\pm0.15)\%$
and $\Pi_{\bar{\Lambda}}=(1.38\pm0.30)\%$. 
As will be discussed at the end of Sec. \ref{exp}, this implies that the
matter created in ultra-relativistic heavy ion collisions is the most
vortical fluid ever produced in the laboratory. 

In this note, we give a brief overview about recent developments in
theories and experiments on the global and local spin polarization
in heavy ion collisions.

\section{Theoretical models in particle polarization}

In this section we first give a brief introduction to the global orbital angular momentum and local vorticity, 
which are basic concepts in this topic. Then we introduce three theoretical models which 
have been widely used in this field. All these models address the same global polarization problem 
in different angles and are consistent to each other. The spin-orbital coupling model is a microscopic model 
and the Wigner function and statistical-hydro model are macroscopic models and of statistical type. 
The thermal average of the local orbital angular momentum in the microscopic model 
gives the vorticity of the fluid in macroscopic models. The same freeze-out formula for the polarization of 
fermions are abtained from the Wigner function and statistical-hydro model, which has been used to 
calulate observables in experiments. 
The Wigner function model is a quantum kinetic approach where quantum effects like 
the chiral magnetic and vortical effect and chiral anomaly can be naturally incorporated. 
The statistical-hydro model is a generalization of the statistical model 
for a thermal system without rotation to a hydrodynamical one with rotation. 
With the statistical-hydro model one can easily derive the spin-vorticity coupling term 
for a system of massive fermions and then the spin polarization density 
which is proportional to vorticity.

\subsection{Global orbital angluar momentum and local vorticity}

Let us consider two colliding nuclei with the beam momentum per nucleon
$\mathbf{p}_{\mathrm{beam}}\equiv p_{\mathrm{beam}}\mathbf{e}_{z}$
(projectile) and $-\mathbf{p}_{\mathrm{beam}}$ (target). The impact
parameter $\mathbf{b}\equiv b\mathbf{e}_{x}$ whose modulus is the
transverse distance between the centers of the projectile and target
nucleus points from the target to the projectile. The normal direction
of the reaction plane or the direction of the global angular momentum
is along $\hat{\mathbf{b}}\times\hat{\mathbf{p}}_{\mathrm{beam}}=-\mathbf{e}_{y}$. 
We should keep in mind that due to event-by-event fluctuations of the nucleon positions, 
the global orbital angular momentum does not in general point to $-\mathbf{e}_{y}$. 
The discussion in this subsection is the ideal case only for theoretical simplicity.  
The magnitude of the total orbital angular momentum $L_{y}$ and the
resulting longitudinal fluid shear can be estimated within the wounded
nucleon model of particle production \cite{Liang:2004ph,Gao2008}.
The transverse distributions (integrated over y) of participant nucleons
in each nucleus can be written as 
\begin{equation}
\frac{dN_{\mathrm{part}}^{\mathrm{P,T}}}{dx}=\int dydz\rho_{A}^{\mathrm{P,T}}(x,y,z,b),
\end{equation}
where $\rho_{A}^{\mathrm{P,T}}$ denotes the number of participant
nucleons in the projectile and target, respectively. One can use models
to estimate $\rho_{A}^{\mathrm{P,T}}$ such as the hard-sphere or
Woods-Saxon model. Then we obtain 
\begin{equation}
L_{y}=-p_{\mathrm{in}}\int dxx\left(\frac{dN_{\mathrm{part}}^{\mathrm{P}}}{dx}-\frac{dN_{\mathrm{part}}^{\mathrm{P}}}{dx}\right).
\end{equation}
The average collective longitudinal momentum per parton can be estimated
as 
\begin{equation}
p_{z}(x,b;\sqrt{s})=p_{0}\frac{dN_{\mathrm{part}}^{\mathrm{P}}/dx-dN_{\mathrm{part}}^{\mathrm{T}}/dx}{dN_{\mathrm{part}}^{\mathrm{P}}/dx+dN_{\mathrm{part}}^{\mathrm{T}}/dx},
\end{equation}
where $p_{0}=\sqrt{s}/[2c(s)]$ denotes the maximum average longtitudinal
momentum per parton. The average relative orbital angular momentum
for two colliding partons separated by $\Delta x$ in the transverse
direction is then $l_{y}\equiv-(\Delta x)^{2}dp_{z}/dx$. 
Note that $l_y$ is expected to be proportional to the local vorticity.

As we all know the strongly coupled quark gluon plasma (sQGP)
can be well described by relativistic hydrodynamic models. So the
sQGP can be treated as a fluid which is characterized by local quantities
such as the momentum, energy and particle-number densities $\mathbf{p}(\mathbf{r})$,
$\epsilon(\mathbf{r})$ and $n(\mathbf{r})$, respectively. The total
angluar momentum of a fluid can be written as $\mathbf{L}=\int d^{3}r\,\mathbf{r}\times\mathbf{p}(\mathbf{r})$.
The fluid velocity is defined by $\mathbf{v}(\mathbf{r})=\mathbf{p}(\mathbf{r})/\epsilon(\mathbf{r})$.
In non-relativistic theory, the fluid vorticity is defined by $\boldsymbol{\omega}=\frac{1}{2}\nabla\times\mathbf{v}(\mathbf{r})$. 
Note that a 1/2 prefactor is introduced in the definition of the vorticity, 
which is different from normal convention, 
this is to be consistent to the convention of 
the vorticity four-vector in relativistic theory. 
For a rigid-body rotation with a constant angular velocity $\bar{\boldsymbol{\omega}}$,
the velocity of a point on the rigid body is given by $v=\bar{\boldsymbol{\omega}}\times\mathbf{r}$.
We can verify that $\boldsymbol{\omega}=\frac{1}{2}\nabla\times(\bar{\boldsymbol{\omega}}\times\mathbf{r})=\bar{\boldsymbol{\omega}}$,
i.e. for a rigid body in rotation the vorticity is identical to the
angular momentum. With the local vorticity, the total angluar momentum
can be re-written as $\mathbf{L}=\int d^{3}r\,\epsilon(\mathbf{r})[r^{2}\boldsymbol{\omega}-(\boldsymbol{\omega}\cdot\mathbf{r})\mathbf{r}]$.
We see that $\mathbf{L}$ is an integral of the moment of inertia
density and the local vorticity. The time evolution of the local velocity
and vorticity field can be simulated through the hydrodynamic model
\cite{Csernai:2013bqa,Csernai:2014ywa,Pang:2016igs}, the AMPT model
\cite{Jiang:2016woz,Li:2017slc} or the HIJING model with a smearing
technique \cite{Deng:2016gyh}.

\subsection{Spin-orbital coupling model}

We first consider a simple model for a spin-1/2 quark scattered in
a static Yukawa potential $V(\mathbf{r})=e^{-m_{D}|\mathbf{r}|}/(4\pi|\mathbf{r}|)$
with $m_{D}$ being the screening mass. The scattering amplitude is
\begin{equation}
\mathcal{M}(p_{\mathrm{i}},\lambda^{\prime}\rightarrow p_{\mathrm{f}},\lambda)=Qu_{\lambda}^{\dagger}(p_{\mathrm{f}})V(\mathbf{q})u_{\lambda^{\prime}}(p_{\mathrm{i}}),
\end{equation}
where $V(\mathbf{q})=1/(\mathbf{q}^{2}+m_{D}^{2})$ is the Fourier
transform of $V(\mathbf{r})$ with $\mathbf{q}=\mathbf{p}_{\mathrm{f}}-\mathbf{p}_{\mathrm{i}}$,
$Q$ denotes the coupling constant, and $u_{\lambda^{\prime}}(p_{\mathrm{i}})$
and $u_{\lambda}(p_{\mathrm{f}})$ are Dirac spinors of the quark
before and after the scattering where $(p_{\mathrm{i}},\lambda^{\prime})$
and $(p_{\mathrm{f}},\lambda)$ are (4-momentum, spin) of the quark
in the initial and final state, respectively. The spin-dependent cross
section can be obtained 
\begin{eqnarray}
\sigma_{\lambda} & = & \frac{1}{2E_{\mathrm{i}}v_{\mathrm{i}}}\frac{1}{2}\sum_{\lambda^{\prime}}\int\frac{d^{3}p_{\mathrm{f}}}{(2\pi)^{3}2E_{\mathrm{f}}}\bigg|\mathcal{M}(p_{\mathrm{i}},\lambda^{\prime}\rightarrow p_{\mathrm{f}},\lambda)\bigg|^{2}(2\pi)\delta(E_{\mathrm{f}}-E_{\mathrm{i}}),
\end{eqnarray}
where $v_{\mathrm{i}}=|\mathbf{p}_{\mathrm{i}}|/E_{\mathrm{i}}$ and
$E_{\mathrm{i}}=\sqrt{\mathbf{p}_{\mathrm{i}}^{2}+m^{2}}$. The polarized
and total cross sections can thus be obtained by $\Delta\sigma=\sigma_{+}-\sigma_{-}$
and $\sigma=\sigma_{+}+\sigma_{-}$. In small angle scatterings, the
corresponding differential cross sections are in the form $d^{2}\sigma/d^{2}\mathbf{x}_{T}\sim K_{0}(m_{D}|\mathbf{x}_{T}|)$
and $d^{2}\Delta\sigma/d^{2}\mathbf{x}_{T}\sim\mathbf{n}\cdot(\mathbf{x}_{T}\times\mathbf{p}_{\mathrm{i}})$,
where $\mathbf{x}_{T}$ is the impact parameter of the scattering
in a small local cell \cite{Liang:2004ph}. We see that the polarized
cross section is proportional to the spin-orbital coupling, $\mathbf{n}\cdot(\mathbf{x}_{T}\times\mathbf{p}_{\mathrm{i}})$,
where $\mathbf{n}$ is the spin quantization direction and $\mathbf{L}=\mathbf{x}_{T}\times\mathbf{p}_{\mathrm{i}}$
is the orbital angular momentum. In order to see the connection of the polarization 
with the spin-ortibal coupling energy $\Delta E_{\mathrm{LS}}$ 
(as in the nuclear shell model), we rewrite the polarization of the particle for small angle scatterings in the static limit ($\mathbf{p}_{\mathrm{i}}\sim 0$), 
\begin{equation}
\Pi\sim\frac{\Delta\sigma}{\sigma}\sim\frac{m_{D}|\mathbf{p}_{\mathrm{i}}|}{E_{\mathrm{i}}(E_{\mathrm{i}}+m)}\sim\frac{m_{D}|\mathbf{p}_{\mathrm{i}}|}{m^{2}}\sim\frac{\Delta E_{\mathrm{LS}}}{E_{0}}
\end{equation}
which $\Delta E_{\mathrm{LS}}$ is given by 
given by 
\begin{equation}
\Delta E_{\mathrm{LS}}\sim\mathbf{L}\cdot\mathbf{S}\frac{1}{m^{2}r}\cdot\frac{dV}{dr}\sim\frac{1}{m^{2}}(E_{0}m_{D}^{2})\frac{|\mathbf{p}_{\mathrm{i}}|}{m_{D}}
\end{equation}
where $E_{0}$ is an energy scale, $L\sim|\mathbf{p}_{\mathrm{i}}|/m_{D}$
is the angular momentum of the particle, $r^{-1}dV/dr\sim E_{0}m_{D}^{2}$
is the potential gradient divided by the typical range of the potential
$r\sim1/m_{D}$.      

One can elaborate the spin-orbital coupling model by considering a
more realistic quark-quark scattering at a transverse distance of
$\mathbf{x}_{T}$, whose polarized differential cross section is proportional
to the spin-orbital coupling $\mathbf{n}\cdot(\mathbf{x}_{T}\times\mathbf{p}_{\mathrm{i}})$,
similar to the case of the static potential \cite{Gao2008}.

\subsection{Wigner function method}

As the spin-orbital coupling involves a particle's angular momentum, 
we have to know a particle's position and momentum simultaneously. 
In the classical theory, we use the phase space distribution function of particles, 
while in quantum theory we have to use the Wigner function, 
a quantum analogue of the distribution function. 

In relativistic quantum theory, the spin four-vector of a massive particle 
is defined as the Pauli-Lubanski pseudo-vector , 
$\hat{S}^{\mu}=-\frac{1}{2m}\hat{J}_{\nu\rho}^{\mathrm{S}}\hat{P}_{\sigma}$,
which satisfies $[\hat{S}^{\mu},\hat{P}^{\nu}]=0$, $\hat{S}^{\mu}\hat{P}_{\mu}=0$
and $\hat{S}^{\mu}\hat{S}_{\mu}=-S(S+1)$ with $S$ is spin quantum
number of the particle. For massive fermions with spin 1/2, 
we can express its spin tensor density in terms of the Wigner function \cite{Fang:2016vpj}, 
\begin{eqnarray}
\left\langle M^{\alpha\beta}(x)\right\rangle  & = & \frac{1}{2}\lim_{y\rightarrow0}\mathrm{Tr}\left[\gamma_{0}\sigma^{\alpha\beta}\psi(x-\frac{y}{2})\bar{\psi}(x+\frac{y}{2})\right]\nonumber \\
 & = & \frac{1}{2}\int d^{4}p\mathrm{Tr}\left[\gamma_{0}\sigma^{\alpha\beta}W(x,p)\right].
\end{eqnarray}
Then we can define the spin tensor component of the Wigner function
as 
\begin{eqnarray}
\mathscr{M}^{\alpha\beta}(x,p) & \equiv & \frac{1}{2}\mathrm{Tr}\left[\gamma_{0}\sigma^{\alpha\beta}W(x,p)\right]\nonumber \\
 & = & \frac{1}{2}\left[-\epsilon^{0\alpha\beta\rho}\mathscr{A}_{\rho}+ig^{\alpha0}\mathscr{V}^{\beta}-ig^{\beta0}\mathrm{Tr}(\gamma^{\alpha}W)\right],
\end{eqnarray}
If we take $\alpha\beta=ij$ (spatial indices), we have a simple relation
\begin{eqnarray}
\mathscr{M}^{ij}(x,p) & = & \frac{1}{2}\epsilon^{ijk}\mathscr{A}^{k}(x,p),
\end{eqnarray}
where $\epsilon_{ijk}$ is 3-dimensional anti-symmetric tensor. We
see that one can treat the axial vector component as the spin pseudo-vector
phase space density. So the polarization (or spin) pseudo-vector density
(with a factor 1/2) is \cite{Fang:2016vpj} 
\begin{eqnarray}
\Pi^{\mu}(x) & \approx & \frac{1}{2}\int d^{4}p\mathscr{A}^{\mu}(x,p)
\end{eqnarray}
at the non-relativistic limit. To match the spin four-vector (Pauli-Lubanski pseudo-vector) 
in relativistic case, we should multiply a Lorentz factor $E_{p}/m$ as, 
\begin{eqnarray}
\Pi^{\mu}(x) & \approx & \frac{1}{2m}\int d^{4}pE_{p}\mathscr{A}^{\mu}(x,p).
\end{eqnarray}

The axial component of the Wigner function can be solved perturbatively 
in an expansion of powers of space-time derivative $(\partial _\mu)^n$ and field strength $(F_{\mu\nu})^n$, 
whose zeroth and first order solution are 
\begin{eqnarray}
A_{(0)}^{\alpha} & = & m\left[\theta(p_{0})n^{\alpha}(\mathbf{p},\mathbf{n})-\theta(-p_{0})n^{\alpha}(-\mathbf{p},-\mathbf{n})\right]\delta(p^{2}-m^{2})A,\nonumber \\
A_{(1)}^{\alpha}(x,p) & = & -\frac{1}{2}\hbar\tilde{\Omega}^{\alpha\sigma}p_{\sigma}\frac{dV}{d(\beta p_{0})}\delta(p^{2}-m^{2})-Q\hbar\tilde{F}^{\alpha\lambda}p_{\lambda}V\frac{\delta(p^{2}-m^{2})}{p^{2}-m^{2}},\label{eq:a0-a1}
\end{eqnarray}
where $V=f_{+}+f_{-}$ and $A=f_{+}-f_{-}$ with the phase space distribution
$f_{s}$ for the spin state $s=\pm$ being defined by 
\begin{eqnarray}
f_{s}(x,p) & = & \frac{2}{(2\pi)^{3}}\left[\theta(p_{0})f_{{\rm FD}}(p_{0}-\mu_{s})+\theta(-p_{0})f_{{\rm FD}}(-p_{0}+\mu_{s})\right],\label{eq:dist}
\end{eqnarray}
where $p_{0}\equiv p_{\mu}u^{\mu}$ with $u^{\mu}$ being the fluid
velocity, $f_{{\rm FD}}$ is the Fermi-Dirac distribution function,
and $\mu_{s}$ is the chemical potential corresponding to the spin
state $s$. In Eq. (\ref{eq:a0-a1}), the 4-vector of the spin quantization
direction is given by 
\begin{eqnarray}
n^{\mu}(\mathbf{p},\mathbf{n}) & = & \Lambda_{\;\nu}^{\mu}(-\mathbf{v}_{p})n^{\nu}(\mathbf{0},\mathbf{n})=\left(\frac{\mathbf{n}\cdot\mathbf{p}}{m},\mathbf{n}+\frac{(\mathbf{n}\cdot\mathbf{p})\mathbf{p}}{m(m+E_{p})}\right),\label{eq:polar-cmoving}
\end{eqnarray}
where $\Lambda_{\;\nu}^{\mu}(-\mathbf{v}_{p})$ is the Lorentz transformation
with $\mathbf{v}_{p}=\mathbf{p}/E_{p}$ and $n^{\nu}(\mathbf{0},\mathbf{n})=(0,\mathbf{n})$
is the spin quantization direction in the rest frame of the fermion. 

We note that the polarization pseudo-vector density at the zeroth
order is vanishing if $\mu_{s}$ does not depend on the spin $s$.
The polarization density at the first order ($\sim \omega^{\alpha}, B^{\alpha}$) is obtained by integration
over 4-momentum for $A_{(1)}^{\alpha}(x,p)$, 
\begin{eqnarray}
\Pi_{(1)}^{\alpha} & \approx & \frac{1}{2m}\hbar\beta\int\frac{d^{3}p}{(2\pi)^{3}}\left\{ \left[E_{p}\omega^{\alpha}+QB^{\alpha}\right]\frac{e^{\beta(E_{p}-\mu)}}{[e^{\beta(E_{p}-\mu)}+1]^{2}}\right.\nonumber \\
 &  & \left.+\left[E_{p}\omega^{\alpha}-QB^{\alpha}\right]\frac{e^{\beta(E_{p}+\mu)}}{[e^{\beta(E_{p}+\mu)}+1]^{2}}\right\} ,\label{eq:pol-1}
\end{eqnarray}
where $Q>0$ is the fermion's electric charge. The momentum spectra
of the polarization pseudo-vector at the freezout hypersurface can
be obtained 
\begin{eqnarray}
E_{p}\frac{d\Pi^{\alpha}(p)}{d^{3}p} & \approx & \frac{\hbar}{2m}\beta\frac{1}{(2\pi)^{3}}\int d\Sigma_{\lambda}p^{\lambda}\nonumber \\
 &  & \times\left(\tilde{\Omega}^{\alpha\sigma}p_{\sigma}\pm Q\tilde{F}^{\alpha\sigma}u_{\sigma}\right)f_{\mathrm{FD}}^{\pm}(x,p)\left[1-f_{\mathrm{FD}}^{\pm}(x,p)\right],\label{eq:pol-freeze}
\end{eqnarray}
where $f_{\mathrm{FD}}^{\pm}$ are Dermi-Dirac distribution functions
for fermions ($+$) and anti-fermions ($-$), respectively, and $\Sigma_{\lambda}$
denotes the freezeout hypersurface. In Eqs. (\ref{eq:pol-1},\ref{eq:pol-freeze}),
we have used $\tilde{F}^{\rho\lambda}=\frac{1}{2}\epsilon^{\rho\lambda\mu\nu}F_{\mu\nu}$,
$\tilde{\Omega}^{\xi\eta}=\frac{1}{2}\epsilon^{\xi\eta\nu\sigma}\Omega_{\nu\sigma}$
with $\Omega_{\nu\sigma}=\frac{1}{2}(\partial_{\nu}u_{\sigma}-\partial_{\sigma}u_{\nu})$,
where $\epsilon^{\mu\nu\sigma\beta}$ and $\epsilon_{\mu\nu\sigma\beta}$
are anti-symmetric tensors with $\epsilon^{\mu\nu\sigma\beta}=1(-1)$
and $\epsilon_{\mu\nu\sigma\beta}=-1(1)$ for even (odd) permutations
of indices 0123, so we have $\epsilon^{0123}=-\epsilon_{0123}=1$.
Instead of $\Omega_{\nu\sigma}$, $\tilde{\Omega}^{\xi\eta}$, $F_{\mu\nu}$
and $\tilde{F}^{\rho\lambda}$, we will also use the vorticity vector
$\omega^{\rho}=\frac{1}{2}\epsilon^{\rho\sigma\alpha\beta}u_{\sigma}\partial_{\alpha}u_{\beta}$,
the electric field $E^{\mu}=F^{\mu\nu}u_{\nu}$, and the magnetic
field $B^{\mu}=\frac{1}{2}\epsilon^{\mu\nu\lambda\rho}u_{\nu}F_{\lambda\rho}$. 
One can use Eq. (\ref{eq:pol-freeze}) to calculate the polarization 
of spin-1/2 baryons at freezeout hypersurface in heavy ion collisions 
and compare with experiments. 

We note that the above formalism is to describe the polarization of massive fermions 
with spin 1/2 such as massive quarks or octet baryons of $(1/2)^+$. 
For massless fermions for which the spin vector is not well defined 
but with helicity or chirality, the polarization can be caused by 
the chiral magnetic and vortical effect \cite{Kharzeev:2007jp,Fukushima:2008xe,Kharzeev:2015znc}. 

\subsection{Statistical-hydro model}

The polarization of a partical in a locally rotating fluid can be
described by the statistical-hydro model. The derivation of relativistic
hydrodynamics in quantum statistical theory was proposed in late 1970s
\cite{zubarev:1979} and early 1980s \cite{weert:1982} and further
developed by several authors \cite{Becattini:2009wh,Becattini:2012tc,Becattini:2013fla,Becattini:2015nva,Hayata:2015lga}.
With the density operator, one can calculate the energy-momentum tensor
and current as functions of space-time, $T^{\mu\nu}(x)=\mathrm{Tr}\left[\hat{\rho}\hat{T}^{\mu\nu}(x)\right]\equiv\left\langle \hat{T}^{\mu\nu}(x)\right\rangle $
and $j^{\mu}(x)=\mathrm{Tr}\left[\hat{\rho}\hat{j}^{\mu}(x)\right]\equiv\left\langle \hat{j}^{\mu}(x)\right\rangle $.
One can employ the principle of maximum entropy to derive the density
operator at local equilibrium. We then use Lagrange multiplier to
maximize the entropy under the condition of fixed $T^{\mu\nu}(x)$
and $j^{\mu}(x)$, 
\begin{eqnarray}
S & = & \mathrm{Tr}\left(\hat{\rho}\ln\hat{\rho}\right)+\int_{\Sigma(\tau)}d\Sigma_{\mu}\left\{ \left[\left\langle \hat{T}^{\mu\nu}(x)\right\rangle -T^{\mu\nu}(x)\right]\beta_{\nu}(x)\right.\nonumber \\
 &  & \left.-\left[\left\langle \hat{j}^{\mu}(x)\right\rangle -j^{\mu}(x)\right]\zeta(x)\right\} ,
\end{eqnarray}
where $\Sigma_{\mu}=\Sigma n_{\mu}$ is the space like hypersurface
with $n_{\mu}$ being the time-like vector, $\beta_{\nu}=\beta u_{\nu}$
with $u_{\nu}$ being the fluid velocity. in which leads to $\hat{\rho}_{\mathrm{LE}}$
at local equilibrium (LE), 
\begin{equation}
\hat{\rho}_{\mathrm{LE}}=\frac{1}{Z}\exp\left[\int_{\Sigma(\tau)}d\Sigma_{\mu}\left(-T^{\mu\nu}\beta_{\nu}+\zeta\hat{j}^{\mu}\right)\right].
\end{equation}
Given $n_{\mu}$, one can determine the local equilibrium value of
$\beta^{\alpha}$ and $\zeta$ by $n_{\mu}\left\langle \hat{T}^{\mu\nu}(x)\right\rangle _{\mathrm{LE}}=n_{\mu}T^{\mu\nu}(x)$
and $n_{\mu}\left\langle \hat{j}^{\mu}(x)\right\rangle _{\mathrm{LE}}=n_{\mu}j^{\mu}(x)$. 

The global equilibrium of the fluid can be found by imposing the stationary
condition under which the density operator does not depend on a particular
choice of space-like hypersurface $\Sigma$, so we have $\int_{\Sigma_{1}}d\Sigma_{\mu}\hat{\Phi}^{\mu}=\int_{\Sigma_{2}}d\Sigma_{\mu}\hat{\Phi}^{\mu}$,
where $\hat{\Phi}^{\mu}\equiv-\hat{T}^{\mu\nu}\beta_{\nu}+\zeta\hat{j}^{\mu}$,
or in another form 
\begin{equation}
\oint_{\Sigma_{1}+\Sigma_{2}+\Sigma_{T}}d\Sigma_{\mu}\hat{\Phi}^{\mu}=\int_{V}d^{4}x\partial_{\mu}\hat{\Phi}^{\mu}=0,
\end{equation}
where $\Sigma_{T}$ is the transverse surface to $\Sigma_{1}$ and
$\Sigma_{2}$. The above equation leads to 
\begin{eqnarray*}
\partial_{\mu}\hat{\Phi}^{\mu} & = & -\frac{1}{2}\hat{T}^{\mu\nu}(\partial_{\mu}\beta_{\nu}+\partial_{\nu}\beta_{\mu})+(\partial_{\mu}\zeta)\hat{j}^{\mu}=0.
\end{eqnarray*}
So we obtain the stationary conditions 
\begin{equation}
\partial_{\mu}\beta_{\nu}+\partial_{\nu}\beta_{\mu}=0,\;\;\partial_{\mu}\zeta=0,
\end{equation}
where the former condition is called the Killing condition whose solution
is in the form $\beta^{\mu}=\beta u^{\mu}+\varpi^{\mu\nu}x_{\nu}$,
where $\varpi^{\mu\nu}=-\frac{1}{2}(\partial^{\mu}\beta^{\nu}-\partial^{\nu}\beta^{\mu})$.
So we obtain the density operator at global equilibrium 
\begin{equation}
\hat{\rho}_{\mathrm{GE}}=\frac{1}{Z}\exp\left[-\beta u_{\nu}\hat{P}^{\nu}+\frac{1}{2}\hat{J}^{\nu\rho}\varpi_{\nu\rho}+\zeta\hat{Q}\right],
\end{equation}
where $\hat{P}^{\nu}=\int_{\Sigma}d\Sigma_{\mu}\hat{T}^{\mu\nu}$,
$\hat{J}^{\nu\rho}=\int_{\Sigma}d\Sigma_{\mu}(x^{\nu}\hat{T}^{\mu\rho}-x^{\rho}\hat{T}^{\mu\nu})$
and $\hat{Q}=\int_{\Sigma}d\Sigma_{\mu}\hat{j}^{\mu}$. We can also
add the spin tensor to the angular momentum tensor density $\hat{S}^{\mu;\nu\rho}$:
\begin{eqnarray}
\hat{J}^{\nu\rho} & = & \int_{\Sigma}d\Sigma_{\mu}(x^{\nu}\hat{T}^{\mu\rho}-x^{\rho}\hat{T}^{\mu\nu}+\hat{S}^{\mu;\nu\rho})\nonumber \\
 & = & \hat{J}_{\mathrm{OAM}}^{\nu\rho}+\hat{J}_{\mathrm{S}}^{\nu\rho}.
\end{eqnarray}
The spin tensor $\hat{J}_{\mathrm{S}}^{\nu\rho}$ gives the Pauli-Lubanski
pseudo-vector. The expectation value of spin vector is given
by $S^{\mu}=\mathrm{Tr}(\hat{\rho}_{\mathrm{GE}}\hat{S}^{\mu})$.
Then the polarization is obtained by $\Pi^{\mu}=S^{\mu}/S$. 

Since the spin pseudo-vector $\hat{S}^{\mu}$ involves the momentum
operator, we need to know a particle's momentum to evaluate its polarization.
In general, this requires the knowledge of the Wigner function, which
allows to express the mean values of operators as integrals over space-time
and 4-momentum. The mean spin pseudo-vector of a spin-1/2 particle
with 4-momentum $p^{\mu}$, produced at $x^{\mu}$ on particlization
hypersurface, at the leading order in the thermal vorticity reads
\cite{Becattini:2013fla,Becattini:2016gvu} 
\begin{equation}
\Pi^{\mu}(x,p)=-\frac{1}{8m}[1-f_{\mathrm{FD}}(x,p)]\epsilon^{\mu\nu\sigma\rho}p_{\nu}\varpi_{\sigma\rho},\label{eq:beca-pol}
\end{equation}
where $f_{\mathrm{FD}}(x,p)$ is the Fermi-Dirac distribution function.
The mean polarization of the particle with 4-momentum $p^{\mu}$ over
the particlization hypersurface is given by 
\begin{equation}
\Pi^{\mu}(p)=\frac{\int d\Sigma_{\rho}p^{\rho}f_{\mathrm{FD}}(x,p)\Pi^{\mu}(x,p)}{\int d\Sigma_{\rho}p^{\rho}f_{\mathrm{FD}}(x,p)}.\label{eq:beca-pol-free}
\end{equation}
Note that at a constant temperature, Eqs. (\ref{eq:beca-pol},\ref{eq:beca-pol-free})
are consistent to Eq. (\ref{eq:pol-freeze}) \cite{Fang:2016vpj}. 
The parameters in Eq. (\ref{eq:beca-pol-free}) are those in the hydrodynamical models 
which give the temperatures, chemical potentials and fluid velocities 
on the freezeout hypersurface.

\section{Experimental measurements of global polarization}
\label{exp}

The global polarization can be measured by the $\Lambda$ hyperon's
weak decay into a proton and a negatively charged pion. Due to its
nature of weak interaction, the proton is emitted preferentially along
the direction of the $\Lambda$'s spin in the $\Lambda$'s rest frame,
so the parity is broken in the decay process. In this sense, we say
that $\Lambda$ is \textit{self-analyzing} since we can determine
the $\Lambda$'s polarization by measuring the daughter proton's momentum
\cite{Overseth:1967zz}. The solid angle distribution for the daughter
proton in the $\Lambda$'s rest frame is given by 
\begin{eqnarray*}
\frac{dN}{d\Omega^{*}} & = & \frac{1}{4\pi}\left(1+\alpha_{H}\hat{\mathbf{p}}_{\mathrm{p}}^{*}\cdot\boldsymbol{\Pi}_{\Lambda}\right)=\frac{1}{4\pi}\left(1+\alpha_{H}\Pi_{\Lambda}\cos\theta^{*}\right),
\end{eqnarray*}
where $\hat{\mathbf{p}}_{\mathrm{p}}^{*}$ is the direction of the
daughter proton's momentum in the $\Lambda$'s rest frame, $\boldsymbol{\Pi}_{\Lambda}$
is the $\Lambda$'s polarization vector with its modulus $\Pi_{\Lambda}<1$,
$\theta^{*}$ is the angle between the momentum of the daughter proton's
and that of $\Lambda$, and $\alpha_{H}=0.642\pm0.013$ is the $\Lambda$'s
decay constant measured in experiments. The $\Lambda$'s polarization
can be determined by an event average of the proton's momentum direction
in the $\Lambda$'s rest frame, 
\begin{equation}
\Pi_{\Lambda}=\frac{3}{\alpha_{H}}\left\langle \cos\theta^{*}\right\rangle _{\mathrm{ev}}.
\end{equation}

We assume the beam direction is along $\mathbf{e}_{z}$, $\hat{\mathbf{p}}_{\mathrm{beam}}=(0,0,1)$,
and the direction of the impact parameter is $\hat{\mathbf{b}}=(\cos\psi_{\mathrm{RP}},\sin\psi_{\mathrm{RP}},0)$
where $\psi_{\mathrm{RP}}$ is the azimuthal angle of the reaction
plane. The global polarization $\mathbf{L}$ is along $\hat{\mathbf{b}}\times\hat{\mathbf{p}}_{\mathrm{beam}}=(\sin\psi_{\mathrm{RP}},-\cos\psi_{\mathrm{RP}},0)$.
The direction of the daughter proton's momentum in the $\Lambda$'s
rest frame is assumed to be $\hat{\mathbf{p}}_{\mathrm{p}}^{*}=(\sin\theta_{\mathrm{p}}^{*}\cos\phi_{\mathrm{p}}^{*},\sin\theta_{\mathrm{p}}^{*}\sin\phi_{\mathrm{p}}^{*},\cos\theta_{\mathrm{p}}^{*})$.
If $\boldsymbol{\Pi}_{\Lambda}$ is in the direction of the global
polarization $\mathbf{L}$, we have 
\begin{equation}
\cos\theta^{*}=\hat{\mathbf{p}}_{\mathrm{p}}^{*}\cdot\hat{\boldsymbol{\Pi}}_{\Lambda}=\sin\theta_{\mathrm{p}}^{*}\sin(\psi_{\mathrm{RP}}-\phi_{\mathrm{p}}^{*}).
\end{equation}
We can obtain the proton's distribution in $\phi_{\mathrm{p}}^{*}$
after an integration over $\theta_{\mathrm{p}}^{*}$, 
\begin{eqnarray}
\frac{dN}{d\phi_{\mathrm{p}}^{*}} & = & \int_{0}^{\pi}d\theta_{\mathrm{p}}^{*}\sin\theta_{\mathrm{p}}^{*}\frac{dN}{d\Omega^{*}}\nonumber \\
 & = & \frac{1}{8}+\frac{1}{8}\alpha_{H}\Pi_{\Lambda}\sin(\psi_{\mathrm{RP}}-\phi_{\mathrm{p}}^{*}).
\end{eqnarray}
Then we obtain $\Pi_{\Lambda}$ by taking an event average of $\sin(\psi_{\mathrm{RP}}-\phi_{\mathrm{p}}^{*})$
\cite{Abelev:2007zk}, 
\begin{equation}
\Pi_{\Lambda}=-\frac{8}{\pi\alpha_{H}}\left\langle \sin(\phi_{\mathrm{p}}^{*}-\psi_{\mathrm{RP}})\right\rangle _{\mathrm{ev}}.\label{eq:pol-lambda}
\end{equation}
The above equation is similar to that used in directed flow measurements
\cite{Barrette:1996rs,Alt:2003ab,Adams:2005ca}, which allows us
to use the corresponding anisotropic flow measurement technique \cite{Voloshin:1994mz,Poskanzer:1998yz}.
The reaction plane angle in Eq. (\ref{eq:pol-lambda}) is estimated
by calculating the angle of the first order event plane, so we need
to correct the final results by the reaction plane resolution $R_{\mathrm{EP}}^{(1)}$.
Then we can rewrite Eq. (\ref{eq:pol-lambda}) in terms of the first-order
event plane angle $\Psi_{\mathrm{EP}}^{(1)}$ and its resolution $R_{\mathrm{EP}}^{(1)}$
\cite{Abelev:2007zk}, 
\begin{equation}
\Pi_{\Lambda}=-\frac{8}{\pi\alpha_{H}R_{\mathrm{EP}}^{(1)}}\left\langle \sin\left(\phi_{\mathrm{p}}^{*}-\Psi_{\mathrm{EP}}^{(1)}\right)\right\rangle _{\mathrm{ev}}.\label{eq:pol-lambda-1}
\end{equation}
The first-order event plane angle is estimated experimentally by measuring
the sidewards deflection of the forward- and backward-going fragments
and particles in the STAR's BBC detectors. 

\begin{figure}
\caption{\label{fig:STAR}STAR results for the global $\Lambda$ polarization \cite{STAR:2017ckg}. }
\begin{center}\includegraphics[scale=0.4]{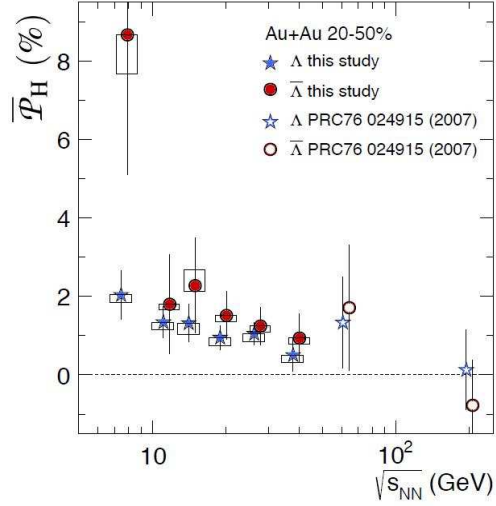}\end{center}
\end{figure}

The STAR's recent measurements for the global $\Lambda$ polarization
at all collisional energies in the Beam Energy Scan (BES) program
are shown in Fig. \ref{fig:STAR} \cite{STAR:2017ckg}. At each energy, a positive polarization
at the level of $(1.1-3.6)\sigma$ is observed for $\Lambda$ and
$\bar{\Lambda}$. Taking all data at different energies into account,
the global polarization for $\Lambda$ and $\bar{\Lambda}$ are $\Pi_{\Lambda}=(1.08\pm0.15)\%$
and $\Pi_{\bar{\Lambda}}=(1.38\pm0.30)\%$ respectively. 
Although the experimental uncertainties are too large to state so with confidence, 
there may be some indication for anti-Lambdas to have larger polarization than Lambdas. 
Such a difference could in principle be caused by magnetic coupling of their opposite magnetic moments to the magnetic field. However, a quick calculation shows 
that even for the largest magnetic fields that could be expected in these collisions 
the effect of this spin-magnetic coupling on the polarization signal would be 
at most a small fraction of a percent and thus invisible in this experiment.
Another source of difference may possibly be due to more Pauli blocking effect for
fermions than anti-fermions in lower collisional energies where fermions
have non-vanishing chemical potentials \cite{Fang:2016vpj,Aristova:2016wxe}.
But still such a difference is too small to be observed given 
the present experimental error bars \cite{STAR:2017ckg,Becattini:2016gvu}. 
The global polarization decreases with increasing collision energy. This is consistent with the observation that longitudinal boost-invariance for the longitudinal expansion becomes a better approximation at higher energies \cite{Li:2017slc,Karpenko:2016jyx}, and that a boost-invariant longitudinal flow profile has a vanishing vorticity component orthogonal to the reaction plane.

The fluid vorticity can be estimated from the data by the hydro-statistical
model $\omega\approx T(\Pi_{\Lambda}+\Pi_{\bar{\Lambda}})$, where
$T$ is the temperature of the fluid at the moment of particle freezeout.
The polarization data averaged over collisional energies imply that
the vorticity is about $(9\pm1)\times10^{21}\,\mathrm{s}^{-1}$. This
is much larger than any other fluids that exist in the universe. Then
the sQGP created in heavy ion collisions is not only the hottest,
least viscous, but also the most vortical fluid that is ever produced
in the laboratory. 

\textit{Acknowledgment}. QW thanks M. Lisa and F. Becattini for helpful
discussions. QW is supported in part by the Major State Basic Research
Development Program (MSBRD) in China under the Grant No. 2015CB856902
and 2014CB845402 and by the National Natural Science Foundation of
China (NSFC) under the Grant No. 11535012. 

\bibliographystyle{elsarticle-num}
\bibliography{ref-1}

\end{document}